# Casimir–Polder force out of thermal equilibrium

G.V. Dedkov, A.A. Kyasov
*Nanoscale Physics Group, Kabardino –Balkarian State University, Nalchik, 360004, Russian Federation*
E-mail: gv_dedkov@mail.ru

We calculate equilibrium and nonequilibrium Casimir –Polder force felt by a small particle (an atom) near a flat substrate using the framework of fluctuation electrodynamics in configuration of a small sphere above a flat substrate. A numerical example is given for a SiC micron –sized particle above a SiC substrate. Different temperature configurations are considered.

PACS numbers : 34.35.+a; 12.20.-m; 37.10 Vz

### 1. Introduction

At zero temperature, the fluctuation electromagnetic force felt by an atom near an ideally conducting metal plate was first pioneered by Casimir and Polder [1]. In broader sense, the Casimir –Polder (CP) force is usually referred to a conservative force between a bulk object with arbitrary dielectric properties and a gas –phase atom. Lifshitz [2] developed general theory of fluctuation electromagnetic forces between two dielectric semispaces at a temperature $T$ in thermal equilibrium with plane parallel boundary surfaces separated by a gap of width much larger than interatomic spacing. In what follows this geometrical configuration will be referred to as configuration "1". If the material of one of the plates is rarified, the general Lifshitz formula allows to get the equilibrium atom –wall force using the limit $\varepsilon_2 - 1 \to 4\pi n\alpha$, where $\varepsilon_2$, $n$ and $\alpha$ are the dielectric permittivity, density and atomic polarizability of rarified plate material, respectively [3,4]. Another (primary) geometrical configuration will be referred to a small particle near a flat substrate (configuration "2"). Recently, a comprehensive review of basic results obtained in this configuration has been given in our papers [5]. Configuration 2 turns out to be useful not only in the problem of conservative CP force. It allows also to study dynamical effects of the CP force for the ground- and excited –state atoms [6,7], dissipative tangential forces and radiation heat transfer between moving particles and substrates [5,8,9].

Much less studied are the situations out of thermal equilibrium, when different parts of the system have different temperatures, being locally in thermal equilibrium (the global system is out of equilibrium but in a stationary regime). Speaking most generally, in configuration 2 one needs to consider a particle with temperature $T_1$ embedded in vacuum background with temperature $T_3$ near a dielectric substrate with temperature $T_2$. These studies are motivated by various technological applications [10,11] and fundamental aspects [12].

Regarding the nonequilbrium CP force, several important results have been obtained to date. So, Henkel at. al. [13] has considered the case $T_1 = T_3 = 0, T_2 = T$, i.e. the case of a cold particle in cold vacuum background in the near field of hot substrate. Antezza et. al. [14] has considered the case $T_1 = 0, T_2 \neq T_3$. The first experimental measurement of a temperature dependence of the CP (atom –wall) force has been done in [15]. The authors have noticed a good agreement with the theory [14].

The aim of this paper is twofold. First, we analyze the temperature configuration $T_1 = T_3 = 0, T_2 = T$ in more detail, using a theoretical basis, which has been developed by us for configuration 2 [5]. Second, we consider some other nonequilibrium configurations, which are not covered by the theory [13,14], and discuss relevance of the obtained results with those corresponding to equilibrium configurations $T_1 = T_2 = T_3 = 0$, $T_1 = T_2 = T_3 = T$

For a numerical example we calculate equilibrium and nonequilibrium CP forces for a micron sized *SiC* particle above a *SiC* substrate.

## 2. Theoretical background

Our approach is based on the expression for the averaged Lorentz force applied to a small neutral particle moving in fluctuating electromagnetic field generated by surrounding bodies. Specifically, the used coordinate system and geometrical arrangement corresponding to configuration 2 are described in [5]. Within the dipole approximation of the fluctuation electromagnetic theory, the averaged Lorentz force corresponding to the particle – surface interaction is shown to be conveniently cast in the form [5]

$$\mathbf{F} = \langle \nabla(\mathbf{dE} + \mathbf{mB}) \rangle \qquad (1)$$

where **d** and **m** are the particle fluctuation electric and magnetic dipole moments, **E** and **B** are the components of fluctuating electromagnetic field. Each of them consists of spontaneous and induced terms, and angular brackets denote total statistical averaging. A projection **F** onto direction *z* of the coordinate system, normal to the surface, corresponds to the Casimir force. Eq.(1) can be adequately applied to both resting and moving particles embedded in fluctuating electromagnetic field. The second term of Eq.(1) takes into account magnetic polarization of a particle which is assumed to be non magnetic in its rest frame. Magnetic polarization proves to be very important in vacuum contacts of metallic bodies [16,17]. At full thermodynamic equilibrium, the particle, substrate and vacuum background are assumed to have equal temperature $T$. In partly nonequilibrium case, when the resting particle has temperature $T_1$, while the substrate and vacuum background are in equilibrium at temperature $T_2$, the resulting Casimir force is given by (see [5] and references therein for more details)

$$F_z = -\frac{2\hbar}{\pi^2} \iiint_{k>\omega/c} d\omega\, d^2k \exp(-2q_0 z) \cdot$$
$$\cdot \left\{ \begin{array}{l} \coth\dfrac{\hbar\omega}{2k_B T_1}\left[ \operatorname{Re} R_e(\omega,\mathbf{k})\alpha_e''(\omega) + \operatorname{Re} R_m(\omega,\mathbf{k})\alpha_m''(\omega) \right] + \\ +\coth\dfrac{\hbar\omega}{2k_B T_2}\left[ \operatorname{Im} R_e(\omega,\mathbf{k})\alpha_e'(\omega) + \operatorname{Im} R_m(\omega,\mathbf{k})\alpha_m'(\omega) \right] \end{array} \right\} -$$
$$-\frac{2\hbar}{\pi^2} \iiint_{k<\omega/c} d\omega\, d^2k \cos(2\tilde{q}_0 z) \cdot \left\{ R_e, R_m \to \tilde{R}_e, \tilde{R}_m \right\} -$$
$$-\frac{2\hbar}{\pi^2} \iiint_{k<\omega/c} d\omega\, d^2k \sin(-2\tilde{q}_0 z) \cdot$$
$$\cdot \left\{ \begin{array}{l} \coth\dfrac{\hbar\omega}{2k_B T_1}\left[ \operatorname{Im} \tilde{R}_e(\omega,\mathbf{k})\alpha_e''(\omega) + \operatorname{Im} \tilde{R}_m(\omega,\mathbf{k})\alpha_m''(\omega) \right] - \\ -\coth\dfrac{\hbar\omega}{2k_B T_2}\left[ \operatorname{Re} \tilde{R}_e(\omega,\mathbf{k})\alpha_e'(\omega) + \operatorname{Re} \tilde{R}_m(\omega,\mathbf{k})\alpha_m'(\omega) \right] \end{array} \right\} \quad (2)$$

where the incoming quantities are defined by

$$q_0 = \left(k^2 - \omega^2/c^2\right)^{1/2},\ \tilde{q}_0 = \left(\omega^2/c^2 - k^2\right)^{1/2},\ k^2 = k_x^2 + k_y^2,\ \mathbf{k} = (k_x, k_y)$$
$$q = \left(k^2 - \varepsilon(\omega)\mu(\omega)\omega^2/c^2\right)^{1/2},\ \tilde{q} = \left(\varepsilon(\omega)\mu(\omega)\omega^2/c^2 - k^2\right)^{1/2} \quad (3)$$

$$\Delta_e(\omega) = \left(\frac{\varepsilon(\omega)q_0 - q}{\varepsilon(\omega)q_0 + q}\right),\ \tilde{\Delta}_e(\omega) = \left(\frac{\varepsilon(\omega)\tilde{q}_0 - \tilde{q}}{\varepsilon(\omega)\tilde{q}_0 + \tilde{q}}\right) \quad (4)$$

$$\Delta_m(\omega) = \left(\frac{\mu(\omega)q_0 - q}{\mu(\omega)q_0 + q}\right),\ \tilde{\Delta}_m(\omega) = \left(\frac{\mu(\omega)\tilde{q}_0 - \tilde{q}}{\mu(\omega)\tilde{q}_0 + \tilde{q}}\right) \quad (5)$$

$$R_e(\omega,k) = (2k^2 - \omega^2/c^2)\Delta_e(\omega) + (\omega^2/c^2)\Delta_m(\omega) \quad (6)$$

$$R_m(\omega,k) = (2k^2 - \omega^2/c^2)\Delta_m(\omega) + (\omega^2/c^2)\Delta_e(\omega) \quad (7)$$

$$\tilde{R}_e(\omega,k) = (2k^2 - \omega^2/c^2)\tilde{\Delta}_e(\omega) + (\omega^2/c^2)\tilde{\Delta}_m(\omega) \quad (8)$$

$$\tilde{R}_m(\omega,k) = (2k^2 - \omega^2/c^2)\tilde{\Delta}_m(\omega) + (\omega^2/c^2)\tilde{\Delta}_e(\omega) \quad (9)$$

Moreover, $\alpha_{e,m}(\omega)$ denotes the frequency dependent dielectric and magnetic polarizabilities of the particle, one –primed and double –primed are the corresponding real and imaginary parts, $\varepsilon(\omega)$ and $\mu(\omega)$ are the dielectric and magnetic permittivities of the plate material (for simplicity, in the following we assume $\mu(\omega) = 1$). Integration over the two–dimensional wave vector in Eq. (2) is performed in the first coordinate quadrant, and over the frequencies –in the domain $0 < \omega < \infty$. The structure of the expression $\{R_e, R_m \to \tilde{R}_e, \tilde{R}_m\}$ is identical to that one in the first integrand term of Eq. (2) with account of the above replacement. Despite that Eq.(2) does not depend on the background temperature $T_3$, the latent assumption $T_2 = T_3$ is essentially implied.

At full equilibrium, $T_1 = T_2 = T_3 = T$, using a representation of Matsubara frequencies, Eq.(2) reduces to the well–known form, in which the integrals over the domains $k > \omega/c$ and $k < \omega/c$ are combined into a single integral over the wave vector $k$, coupled with the sum over the imaginary frequencies $i\xi_n$,

$$F^{eq}(z,T) = -2k_B T \sum_{n=0}^{\infty} a_n \int_0^{\infty} dk k [R_e(i\xi_n,k)\alpha_e(i\xi_n) + R_m(i\xi_n,k)\alpha_m(i\xi_n)] \exp\left(-2\sqrt{k^2 + \xi_n^2/c^2}\,z\right) \quad (10)$$

where $\xi_n = 2\pi k_B T n/\hbar, a_n = 1 - \delta_{0n}/2$.
Provided that $\alpha_m(\omega) = 0$, Eq. (10) explicitly coincides with the results [4], while at a zero temperature $T = 0$ it takes the form

$$F_0(z) = -\frac{\hbar}{\pi} \int_0^{\infty} d\xi \int_0^{\infty} dk k \exp(-2\sqrt{k^2 + \xi^2/c^2}\,z)[R_e(i\xi,k)\alpha_e(i\xi) + R_m(i\xi,k)\alpha_m(i\xi)] \quad (11)$$

In the case out of equilibrium, at $T_1 = 0, T_2 = T_3 = T$, as it follows from (2), the Casimir force turns out to be

$$F^{neq}(0,T,T,z) = F^{eq}(z,T) + \Delta F(z,T) \quad (12)$$

$$\Delta F(z,T) = \frac{2\hbar}{\pi} \int_0^{\infty} d\omega \, \Pi(\omega,T) \alpha_e''(\omega) \operatorname{Re}\left\{\int_0^{\infty} dk k [\exp(-2q_0 z) R_e(\omega,k)]\right\} + \{\alpha_e'' \to \alpha_m'', R_e \to R_m\} \quad (13)$$

where $\Pi(\omega,T) = (\exp(\hbar\omega/k_B T) - 1)^{-1}$. Note that Eqs.(12),(13) can not be obtained using configuration 1 and the limit of rarified medium. If one needs to separate the cold and thermal parts of the CP force (12) from each other, another form of this expression seems to be more preferable. It can be obtained by making use of the identity $\coth(x/2) = 1 + 2/(\exp(x) - 1)$ in Eq.(2). Then, after evident transformations we get

$$F^{neq}(0,T,T,z) = F_0(z) + F_{th}^{neq}(0,T,T,z) \quad (14)$$

where the function $F_0(z)$ is determined by Eq.(11), and the thermal contribution $F_{th}^{neq}(0,T,T,z)$ is given by (for the sake of simplicity, here and in the following we put $\alpha_m(\omega) = 0$)

$$F_{th}^{neq}(0,T,T,z) = -\frac{2\hbar}{\pi} \int_0^{\infty} d\omega \, \alpha_e'(\omega) \Pi(\omega,T) \left\{ \int_{\omega/c}^{\infty} dk k \operatorname{Im} R_e \exp(-2q_0 z) + \int_0^{\omega/c} dk k \operatorname{Im}[\tilde{R}_e \exp(2i\tilde{q}_0 z)] \right\} \quad (15)$$

The first and second terms in Eq.(15) represent the contributions from evanescent and propagating (traveling) modes. The traveling mode contribution is generated due to the thermal equilibrium between vacuum background and substrate, resulting in interference of the traveling modes emitted by the substrate and those coming from vacuum and impinging the substrate surface. In the case $T_2 \neq T_3$, the last term in Eq. (15) must be modified. On the contrary, the first term (15) depends only on the substrate temperature and does not depend on the environment temperature. A difference between the situations $T_2 = T_3 = T$ and $T_2 \neq T_3$ is due to the different structure of fluctuation electromagnetic field above the substrate. In the last case the radiation flux between the substrate and vacuum space results in wind force on a particle, being independent on distance $z$ [13], while in the former one the flux of radiation and the wind force are absent.

Now let us consider the temperature configuration corresponding to a particle with temperature $T_1$, a substrate with temperature $T_2$ and cold environment with $T_3 = 0$. Then, according to Levine and Rytov [18], the spectral energy density of fluctuating electromagnetic field above the substrate surface is given by

$$\rho(\omega) = \frac{\hbar}{4\pi^2} \coth\left(\frac{\hbar \omega}{2k_B T_2}\right) \cdot \left\{ \frac{\omega^2}{2c^2} \int_0^{\omega/c} \frac{dkk}{|q_0|} \left(2 - |\Delta_e|^2 - |\Delta_m|^2\right) + \int_{\omega/c}^{\infty} dk k^3 \operatorname{Im}\left[\frac{\exp(-2q_0 z)}{q_0}(\Delta_e + \Delta_m)\right] \right\} \qquad (16)$$

Comparing Eq.(16) with analogous one corresponding to the equilibrium case $T_2 = T_3 = T$ [19,20], we can make sure that the difference between them is related only with the first term (16), corresponding to the contribution of propagating modes. The contribution from evanescent modes of the surface (the second term in Eq.(16)) does not change. Bearing in mind this feature allows one to understand our final result for the CP force in the temperature configuration $(T_1; T_2; 0)$:

$$F_z = F_z^{(S)} + F_z^{(R)} \qquad (17)$$

$$F_z^{(S)} = -\frac{\hbar}{\pi}\int_0^\infty d\xi \int_0^\infty dk k \exp(-2\sqrt{k^2+\xi^2/c^2}z)R_e(i\xi,k)\alpha_e(i\xi) -$$

$$-\frac{2\hbar}{\pi}\int_0^\infty d\omega \operatorname{Re}\alpha_e(\omega)\Pi(\omega,T_2)\int_{\omega/c}^\infty dk k \operatorname{Im}R_e(\omega,k)\exp(-2q_0 z) - \quad (18)$$

$$-\frac{2\hbar}{\pi}\int_0^\infty d\omega \operatorname{Im}\alpha_e(\omega)\Pi(\omega,T_1)\left\{\begin{array}{l}\int_{\omega/c}^\infty dk k \operatorname{Re}[R_e(\omega,k)\exp(-2q_0 z)]+ \\ +\int_0^{\omega/c} dk k \operatorname{Re}[\tilde{R}_e(\omega,k)\exp(2i\tilde{q}_0 z)]\end{array}\right\}$$

$$F_z^{(R)} = \frac{1}{c}\int_0^\infty d\omega S_z(\omega)\left\{\sigma_a(\omega) + \int_0^\pi (1-\cos\theta)\frac{d\sigma_s}{d\Omega}d\Omega\right\} \quad (19)$$

where $F_z^{(R)}$ is the wind force generated by absorption and scattering of the substrate thermal radiation on the particle. Eq (19) is valid at $R \ll 2\pi\hbar c/k_B T_2$ [21]. For spheres of larger radius one should use the general Mie theory [22]. However, it should be noticed that Eq.(2) and all other results are valid at $R/z \ll 1$. The spectral radiation flux of the substrate (see Eq.(19)) is given by

$$S_z(\omega) = \frac{\hbar\omega}{4\pi^2}\Pi(\omega,T)\int_0^{\omega/c} dk k\left(2-|\Delta_e|^2-|\Delta_m|^2\right) \quad (20)$$

Moreover, $\sigma_a(\omega)$ and $d\sigma_s/d\Omega$ in Eq. (19) denote the particle cross-sections for absorption and scattering of the unpolarized electromagnetic radiation [23]:

$$\sigma_a(\omega) = \frac{4\pi\omega}{c}\operatorname{Im}\alpha_e(\omega) \quad (21)$$

$$\frac{d\sigma_s}{d\Omega} = \frac{\omega^4}{2c^4}|\alpha_e(\omega)|^2(1+\cos^2\theta) \quad (22)$$

With account of (20)-(22), performing a simple integration over the scattering angle $\theta$, Eq.(19) takes the form

$$F_z^{(R)} = \frac{\hbar}{\pi c^2}\int_0^\infty d\omega \omega^2 \Pi(\omega,T_2)\left[\operatorname{Im}\alpha_e(\omega) + \frac{2}{3}(\omega/c)^3|\alpha_e(\omega)|^2\right]\int_0^{\omega/c} dk k\left(2-|\Delta_e|^2-|\Delta_m|^2\right) \quad (23)$$

Eqs.(10)-(15), (18) and (23) are the main results of this paper.

### 3. Numerical examples

Let us first consider nonequilibrium CP force between the ground state atom embedded in cold vacuum background near the substrate with temperature $T_2 = T$. In this case one must take $T_1 = 0$ in Eq.(18). The formula obtained in this way explicitly coincides with that one reported by Antezza et.al. [14]. However, Eq. (17) still includes the radiation wind component (23), which turns out to be important at large particle separations from the surface. For the case atom –wall interaction Eq. (23) can be simplified further using a classical model for the atomic polarizability,

$$\alpha_e(\omega) = \frac{e^2}{m}\frac{1}{\omega_0^2 - \omega^2 - i\gamma\omega} \quad (24)$$

where $e, m, \omega_0$ are the electron charge, mass and frequency of the spectral line, $\gamma = (2e^2/3mc^2)(\omega_0^2/c)$ is the "natural" radiation length. Typically, $k_B T/\hbar\omega_0 \ll 1$ and the dielectric function $\varepsilon(\omega)$ which comes into the reflection factors $|\Delta_e|, |\Delta_m|$ can be replaced by the static value $\varepsilon(0)$. Then, assuming $\varepsilon(0) = 9.4$ (a sapphire substrate) we get from (23)

$$F_z^{(R)} \approx 18.4 k_B T \left(\frac{e^2}{mc^2}\right)^2 \left(\frac{\omega_0}{c}\right)^3 \left(\frac{k_B T}{\hbar\omega_0}\right)^5$$
$$+ 764 k_B T \left(\frac{e^2}{mc^2}\right)^2 \left(\frac{k_B T}{\hbar\omega_0}\right)^4 \left(\frac{k_B T}{\hbar c}\right)^3 \quad (25)$$

The numerical coefficients in Eq.(25) turn out to be slightly dependent on $\varepsilon(0)$. For example, for the $^{87}Rb$ atom ($\omega_0 = 5.35 \cdot 10^{15} s^{-1}$) and a sapphire substrate the repulsive force (25) dominates the attraction force $F_z^{(S)}$ (Eq.(18)) at $z > 9mm$ and $z > 0.7mm$, if the substrate temperature equals $300K$ and $2000K$, respectively.

Furthermore, we have performed a numerical evaluation of the thermal CP force for a micron sized $SiC$ particle above a $SiC$ substrate. Eqs.(10),(11), (15), (18) and (23) allow to discriminate the following temperature configurations:

1) $T_1 = T_2 = T_3 = 0$; 2) $T_1 = T_2 = T_3 = T$;
3) $T_1 = 0, T_2 = T_3 = T$; 4) $T_1 = T_3 = 0, T_2 = T$;
5) $T_1 = T, T_2 = T_3 = 0$.

The dielectric function $\varepsilon(\omega)$ has been approximated by the resonance model [13,24] with the same numerical parameters. The particle radius has been assumed to be $R = 1\mu m$ and the polarizability

$$\alpha_e(\omega) = R^3(\varepsilon(\omega)-1)/(\varepsilon(\omega)+2).$$

Fig.1 compares the numerical results corresponding to the temperature configurations 1-4 (lines 1-4). Line 5 corresponds to the wind force (23). Fig.2 shows the results corresponding to configuration 5 at different temperatures of the particle, comparing them with zero temperature CP force.

### 4. Conclusion

To conclude, we have calculated the CP force in the particle –plate configuration for different nonequilibrium thermal configurations and discuss their relevance to the known results at full equilibrium. We confirm the results obtained by Henkel et.al. [13] and Antezza et.al. [14] in the case of cold particle embedded in cold vacuum background near the heated substrate. We also have obtained the expressions for the CP force in new configurations: "a cold particle in hot environment near a hot substrate" and "a hot particle in cold environment near a cold substrate". Also, the expression for the wind force from heated substrate is derived. The numerical results demonstrate new interesting features of the thermal CP forces at particle –surface separations ranging in the interval from 10 to 50 $\mu m$.

____________________________________

FIGURE CAPTIONS :

Fig.1 Casimir –Polder force between a SiC spherical particle ($R = 1\mu m$) and a SiC substrate at different temperature configurations.

Fig. 2 The same as in Fig.1

FIGURE 1

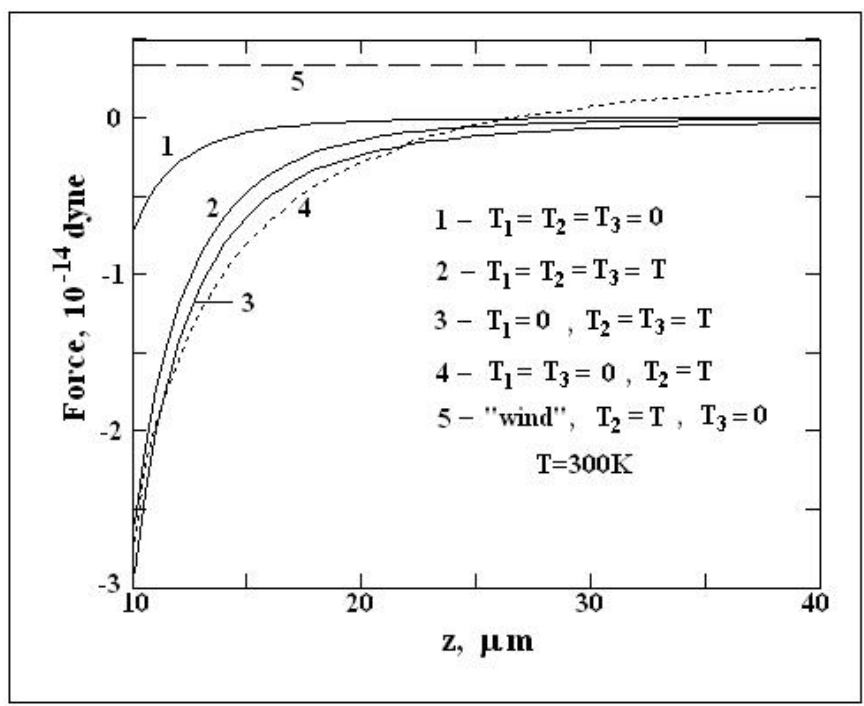

FIGURE 2

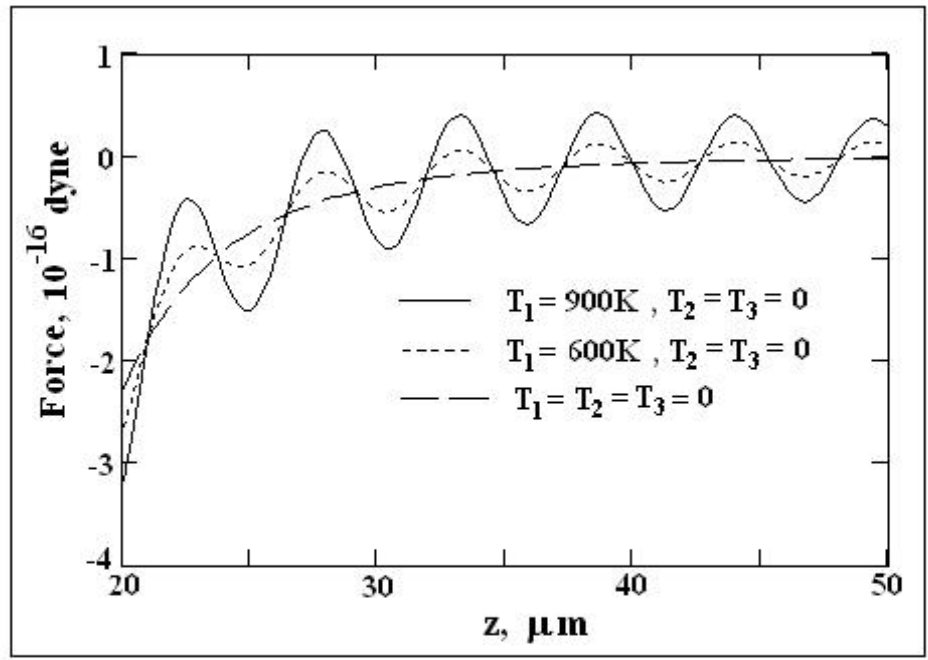